\documentstyle[prl,aps]{revtex}
\input epsf.tex

\newcommand{\nc}{\newcommand}
\nc{\beq}{\begin{equation}}
\nc{\eeq}{\end{equation}}
\nc{\beqa}{\begin{eqnarray}}
\nc{\eeqa}{\end{eqnarray}}
\nc{\lra}{\leftrightarrow}
\def\sfrac#1#2{{\textstyle{#1\over #2}}}
\nc{\sss}{\scriptscriptstyle}
{\nc{\lsim}{\mbox{\raisebox{-.6ex}{~$\stackrel{<}{\sim}$~}}}
{\nc{\gsim}{\mbox{\raisebox{-.6ex}{~$\stackrel{>}{\sim}$~}}}

\def\hp{\hat\phi}
\def\eps{\epsilon}
\def\hmo{\hat m_1}
\def\hmz{\hat m_0}
\def\sqe14{\sqrt{\phantom{iAAA^1}} \!\!\!\!\!\!\!\!\!\!\!\!\!\!\!\!\!\!
   1+\sfrac{4}{\hmo} }

\begin{document}
\twocolumn[\hsize\textwidth\columnwidth\hsize\csname@twocolumnfalse%
\endcsname

\title{Brane-World Cosmology of Modulus Stabilization with a Bulk Scalar Field}

\author{James M.~Cline and Hassan Firouzjahi}

\address{Physics Department, McGill University,\\
3600 University Street, Montr\'eal, Qu\'ebec, Canada H3A 2T8}

\maketitle

\begin{abstract} We point out that the potential of Goldberger and Wise
for stabilizing the distance between two 3-branes, separated from each
other along an extra dimension with a warp factor, has a metastable
minimum when the branes are infinitely separated.  The classical
evolution of the radion (brane separation) will place it in this false
minimum for generic initial conditions.  In particular, inflation could
do this if the expansion rate is sufficiently large.  We present a
simplified version of the Goldberger-Wise mechanism in which the radion
potential can be computed exactly, and we calculate the rate of
thermal transitions to the true minimum, showing that model parameters can be
chosen to ensure that the universe reaches the desired final state.
Finiteness of bulk scalar field brane potentials 
can have an important impact on the nucleation rate,
and it can also significantly increase the predicted mass of the
radion.\\

\vskip0.05in
{PACS: 98.80.Cq \hfill McGill 00-14}
\vskip0.2in
\end{abstract}

]

\section{Introduction}

One of the most striking proposals in current elementary particle
theory is the existence of extra dimensions which are hidden from us,
not necessarily by their smallness, but by our confinement to a
four-dimensional slice (a 3-brane) of the full spacetime \cite{ADD}.
Randall and Sundrum \cite{RS} have produced a verion of this scenario
which is particularly attractive because of its apparent links to deep
theoretical developments: the conformal field theory/5D-anti-de Sitter
correspondence and holography \cite{Verlinde}.  On a more practical
level, their idea provides an explanation for why the Planck scale is
so much greater than the weak scale, independently of supersymmetry.
The cosmological implications of this model have also been the subject of
vigorous study \cite{recent}.

The Randall-Sundrum proposal involves a ``Planck brane'' located at
a position $y=0$ in a single additional dimension, and a second ``TeV
brane'' located at $y=1$, in our conventions.  The extra dimension is
permeated by a negative bulk vacuum energy density, so that the space
between the branes is a slice of anti-de Sitter space.  Solving the
5-D Einstein equations results in the line element
\beqa
\label{metric}
	ds^2 &=& e^{-2\sigma(y)}(dt^2 - d\vec x^2) - b^2\, dy^2;  \\
\label{eq:sigma}
	\sigma(y) &=& kby; \qquad y\in [0,1]
\eeqa
The constant $k$ is related to the 4-D and 5-D Planck masses, 
$M$ and $M_p$ respectively, by 
\beq
	k = {M^3\over M_p^2}
\eeq
where $M_p^{-2} = 8\pi G_{\sss N}$ and the 5-D gravitational action is 
$S = \sfrac12 M^3\int d^{\,5}x \sqrt{-g} R$.  The warp factor $e^{-\sigma(y)}$
determines the physical masses of particles on the TeV brane: even if a
bare mass parameter $m_0$ in the TeV brane Lagrangian is of order $M_p$, 
the physical mass  gets rescaled by
\beqa
	&& m_{\rm phys} = e^{-\bar\sigma} m_0;\qquad 
	\bar\sigma \equiv \sigma(1) = kb,
\eeqa
as can be easily derived starting from a scalar field action written covariantly
in terms of the metric (\ref{metric}),
$S = \frac12 \int d^{\,4}x e^{-4\bar\sigma}(e^{2\bar\sigma} (\partial\phi)^2
- m_0^2\phi^2)$, and rescaling $\phi$ so that the kinetic term becomes canonically
normalized.

In order for $m_0$ to be of order 100 GeV, the combination $kb$ must be
approximately 36, so that $e^{-\bar\sigma}\sim 10^{-16}$.  Yet in the
original proposal, the value of $b$, which is the size of the extra
dimension, was completely undetermined.  It is a modulus with no potential,
which is phenomenologically unacceptable.  For one thing, the particle
associated with 4-D fluctuations of $b$, the radion, would couple to matter
on the TeV brane similarly to gravitons, but more strongly by a factor of
$e^{\bar\sigma}$ \cite{GW3}.  This would lead to a fifth force which could
easily have been detected.  Furthermore, a massless radion leads to problems
with cosmology: our brane-universe would have to have a negative energy
density to expand at the expected rate, assuming that energy densities on the
branes are tuned to give a static extra dimension\cite{CGKT}.  It was shown
in refs.\ \cite{CGRT,KKOP} (see also \cite{Ketc}) that this problem disappears when
the size of the extra dimension is stabilized. 

Radion stabilization is therefore a crucial ingredient of the
Randall-Sundrum idea. Goldberger and Wise have presented an elegant
mechanism for accomplishing this \cite{GW}, using a bulk scalar field.
Self-interactions of the field on the branes forces it to take
nonvanishing VEV's, $v_0$ and $v_1$ respectively, which are generally
different from each other.  The field thus has a gradient in the extra
dimension, and the competition between the gradient and potential
energies gives a preferred value for the size of the extra dimension.  In
other words, a potential for the
radion is generated, which has a nontrivial minimum.  It is easy to get
the correct brane separation using natural values of the parameters in the
model.

Roughly speaking, the radion potential has the form 
\beq
	V(\phi) \cong \lambda \phi^4 \left( \left({\phi\over f}\right)^\eps -
	{v_1\over v_0}\right)^2
\eeq
with a nontrivial minimum at $\phi = f(v_1/v_0)^{1/\eps}$.  However, there is
another minimum at $\phi=0$, which represents an infinitely large extra
dimension.  This could not describe our universe, because then
$e^{-\bar\sigma}$ would be zero, corresponding to vanishing particle
masses on the TeV brane.  In the more exact expression for the potential, 
we will show
that $\phi=0$ is actually a false vacuum---it has higher energy than the
nontrivial minimum.  Nevertheless, is is quite likely that the metastable
state could be reached through classical evolution in the early universe. 
The question then naturally arises whether tunneling or thermal transitions
to the desired state occurs. This is the subject of our study. 

Such detailed questions about the viability of the Goldberger-Wise mechanism
are important because there are few attractive alternatives at the moment. 
Ref.\ \cite{Casimir} recently studied Casimir energies of fields between the
branes as a possible origin of a stabilizing potential.  They found that
stabilization in this way is indeed possible, but that the resulting radion
mass is too small to be phenomenologically consistent if the size of the
extra dimension is that taken to be that dictated by the hierarchy problem. 

In section 2 we derive the Goldberger-Wise potential for the radion in
a slightly simplified model which allows for the exact solution of the
potential.  The classical evolution of the radion field is considered
in section 3, where we show that for generic initial conditions, the
universe reaches a state in which the radion is not stabilized, but
instead the extra dimension is expanding.  This is a metastable state
however, and in section 4 the rate of transitions to the minimum energy
state, with finite brane separation, is computed.  Conclusions are given
in section 5.

\section{Radion Potential}

Let $\psi(y)$ be the bulk scalar field which will be responsible for
stabilizing the radion.  Its action consists of a bulk term plus
interactions confined to the two branes, located at coordinate positions
$y=0$ and $y=1$, respectively.  Using the variable $\sigma$ of eq.\
(\ref{eq:sigma}) in place of $y$, the 4-D effective Lagrangian for a static
$\psi$ configuration can be written as
\beqa
\label{psi_L}
{\cal L} &=& - {k\over 2}\int_{-\bar\sigma}^{\bar\sigma} e^{-4\sigma}
	\left( (\partial_\sigma
	\psi)^2 + \hat m^2 \psi^2\right) d\sigma\nonumber\\ &-&
	m_0 (\psi_0 - v_0)^2 - e^{-4\bar\sigma}m_1 (\psi_1 - v_1)^2,
\eeqa
where $\hat m$ is the dimensionless mass $m/k$,
$\psi_i$ are the values of $\psi$ at the respective branes, and the
orbifold symmetry $\psi(\sigma)=\psi(-\sigma)$ is to be understood.  
In 5-D, the field $\psi$, as well as the VEV's $v_i$ on the two branes,
have dimensions of (mass)$^{3/2}$, while the parameters $m_i$ have
dimensions of mass.  Denoting $\partial_\sigma\psi = \psi'$, the
the Euler-Lagrange equation for $\psi$ is
\beqa
\label{ELeq}
	ke^{-4\sigma}\left(\psi'' \right.&-&\left.4\psi'-\hat m^2\psi\right) = 
	2 m_0(\psi_0-v_0)\delta(\sigma) \nonumber\\
	&+& 2 e^{-4\bar\sigma} m_1(\psi_1-v_1)\delta(\sigma-\bar\sigma).
\eeqa
The general solution has the form
\beqa
\label{soln}
	\psi &=& e^{2\sigma}\left(Ae^{\nu\sigma} + Be^{-\nu\sigma}\right);
	\nonumber\\
	\nu &=& \sqrt{4 + \hat m^2} \equiv 2 + \eps.
\eeqa
To get the correct hierarchy between the Planck and weak scales, it
is necessary to take $\hat m^2$ to be small, hence the 
notation $\eps$.

The brane terms induce boundary conditions specifying the discontinuity
in $\psi'$ at the two branes.  Imposing the orbifold symmetry
$\psi(-\sigma) = \psi(\sigma)$, this implies that 
\beqa
\label{bcs}
	\psi'(0) &=& \hmz(\psi_0-v_0)\nonumber\\
	\psi'(\bar\sigma) &=& -\hmo(\psi_1-v_1),
\eeqa
where we defined $\hat m_i = m_i/k$.
Substituting the solution (\ref{soln}) into (\ref{bcs}) allows us
to solve for the unknown coefficients $A$ and $B$ exactly.  In this
respect, the present model is simpler than that originally given in
ref.\ \cite{GW}.  There the brane potentials were taken to be quartic
functions, so that $A$ and $B$ could only be found in the approximation
where the field values $\psi_i$ were very strongly pinned to their
minimum energy values, $v_i$.  In our model this would occur in the limit
$m_i\to\infty$.  However, we can easily explore the effect of leaving
these parameters finite  since $A$ and $B$ can be determined
exactly.  Let us denote
\beq
	\hp = e^{-kb} = e^{-\bar\sigma},
\eeq
which will be convenient in the following, because $\hp$ is proportional
to the canonically normalized radion field.  Then $A$ and $B$
are given by
\beqa
\label{ABeq}
	A &=& \left(- C_1 \hp^\nu + C_2 \hp^2\right) 
	\hp^\nu / D(\hp) \nonumber\\
	B &=& \left(C_3 \hp^{-\nu} - C_4 \hp^2\right) 
	\hp^\nu / D(\hp)
\eeqa
where
\beqa
\label{Ceq}
	C_1 &=& \hmz v_0 (\hmo-\eps)\nonumber\\
	C_2 &=& \hmo v_1 (\hmz +\eps )\nonumber\\
	C_3 &=& \hmz v_0 (\hmo+ 4+\eps)\nonumber\\
	C_4 &=& \hmo v_1 (\hmz -4-\eps)\nonumber\\
	D(\hp) &=& {\left(C_2 C_3 - C_1 C_4 \hp^{2\nu}\right)\over
	\left(\hmz \hmo v_0 v_1 \right)}.
\eeqa
It can be checked that in the limit $\hat m_i\to \infty$, the field
values on the branes approach $\psi_i\to v_i$.  For finite $\hat m_i$,
the competing effect of minimizing the bulk energy causes departures
from these values, however.

The solution for $\psi$ can be substituted back into the Lagrangian
(\ref{psi_L}) to obtain the effective 4-D potential for the radion,
$V(\hp)$.  However, rather than substituting directly, one can
take advantage of the fact that $\psi$ is a solution to the
Euler-Lagrange equation.  Doing a partial integration and using the
boundary terms in (\ref{ELeq}) gives a simpler expression for
$V(\hp)$.  In the general case where the brane potentials 
are denoted by $V_i(\psi)$, one can easily show that 
\beq
	{\cal L} = -\sum_i e^{-4\sigma_i}\left( V_i(\psi_i) - 
	\sfrac12 \psi_i V'_i(\psi_i) \right)
\eeq
In the present case, we obtain
\beqa
	V(\hp) &=& - {\cal L} \nonumber\\ 
	&=& m_0 v_0 (v_0 - \psi_0) + 
	\hp^4 m_1 v_1 (v_1 - \psi_1) \nonumber\\
	&=& m_0 v_0 (v_0 - (A+B) ) \nonumber\\ &+& 
	\hp^4 m_1 v_1 (v_1 - \hp^{-2}(A \hp^{-\nu}
	+ \hp^\nu B) )
\eeqa

In the following, we will be interested in the situation where $V(\hp)$
has a nontrivial minimum for very small values of $\hp\sim 10^{-16}$,
as needed to address the weak scale hierarchy problem.  It is therefore 
a good approximation to expand $V(\hp)$ near $\hp=0$, keeping
only the terms which are larger than $\hp^{2\nu}$.  This is accomplished
by expanding the denominator $D(\hp)$ in eqs.\ (\ref{ABeq}), after which
the potential can be written in the simple form
\beq
\label{Vexp}
V(\hp) = \Lambda \hp^4 \left( \left(1 + \eps_4
- \eps_1\right)\hp^{2\eps} - 2\eta \left(1+\eps_4
\right) \hp^\eps + \eta^2 \right) 
\eeq
where we introduce the notation
\beq
	\eps_4 = {\eps\over 4};\qquad \eps_0 = {\eps\over\hmz};
	\qquad \eps_1 = {\eps\over\hmo},
\eeq
\beq
	\eta = \left(1+\eps_0\right){v_1\over v_0}
\eeq
and 
\beq
\label{Lambdaeq}
	\Lambda = 4 k v_0^2\, {(1+\eps_4) \left(1 + \eps_0\right)^2
	\over \left(1 + \sfrac{4}{\hmo} + \eps_0\right) }
\eeq
In the following it will be convenient for us to rewrite $V(\hp)$ in 
the form
\beq
\label{eq:pot}
	V(\hp) = \Lambda' \hp^4 (\hp^\eps - c_+)(\hp^\eps - c_-)
\eeq
where $c_\pm$ are given by
\beq
\label{cpmeq}
	c_\pm = \eta\left( \left(1+\eps_4\right)\pm \sqrt{
	\left(1+\eps_4\right)^2 - 
	\left(1 + \eps_4
	- \eps_1\right) } \over \left(1 + \eps_4
	- \eps_1\right) \right)
\eeq
and
$\Lambda' = \Lambda (1+\eps_4-\eps_1)$.

\section{Phenomenology and Early Cosmology of the Model}

The warp factor which determines the hierarchy between the weak and
Planck scales can be found by minimizing the potential (\ref{eq:pot}).
Expanding in $\eps$, it has a global minimum and a local maximum at the
respective values $\hp_+$ and $\hp_-$:
\beqa
\label{extrema}
	\hp_{\pm} &=& \left(v_1\over v_0\right)^{1/\eps}
\!\!\left(\!{1\!+\!\eps_0\over 1\!+\!\eps_4\!-\!\eps_1}\!\right)^{1/\eps}
\!\!\left(1 \pm \sqrt{\eps_1+\eps_4-\sfrac12\eps\eps_1}\right)^{1/\eps}
	\nonumber\\
	&\cong& \left(v_1\over v_0\right)^{1/\eps}\!\!\! \exp\left(
	\pm \sqrt{\sfrac{1}{\eps}(\sfrac{1}{\hmo} + \sfrac14)}
	+ \sfrac{1}{\hmz}\!+\! \sfrac{1}{\hmo}\!-\!\sfrac14\right)
\eeqa
The last approximation holds in the limit of small $\eps$, $\eps_0$
and $\eps_1$; it is not
always an accurate approximation for the parameter values of interest,
so we will use the exact expression in any computations which might be
sensitive to the actual value.  The positions of the zeroes of $V$,
$\hp = c_\pm^{1/\eps}$, are slightly greater than $\hp_{\pm}$, by the
factor $(1+\eps_4)^{1/\eps}\cong e^{1/4}$, as can be seen by comparing
(\ref{extrema}) with (\ref{cpmeq}).  The
large hierarchy of $\hp_+\sim 10^{-16}$ is achieved by taking a
moderately small ratio $v_1/v_0<1$ and raising it to a 
large power,\footnote{An alternative possibility, taking $v_1/v_0>1$ 
and $\eps <0$,
corresponding to a negative squared mass in 
the bulk Lagrangian (\ref{psi_L}), does not work.  Although the negative
$m^2$ does not necessarily lead to any instability,
since the field is prevented from going to infinity by the potentials
on the branes, the radion potential has no nontrivial minimum in this
case.}   
$1/\eps$.  This leads to the mass scale which functions like the cutoff
on the TeV brane,
\beq
\label{MTeVeq}
	\hp_+ M_p \equiv M_{\rm TeV}
\eeq
where $M_{\rm TeV}/(1 {\rm\ TeV})$ is a number of order unity, 
which we will take to be
one of the phenomenological free parameters of the model.  The choice
of $M_{\rm TeV}$ specifies precisely where we want our cutoff scale to be.
 In ref.\ \cite{GW} the exponential corrections to 
$(v_1/v_0)^{1/\eps}$ in {(\ref{extrema}) were not considered; these change somewhat the
values one might choose for $v_1/v_0$ and $\eps$ to get the correct
hierarchy.  The factor $e^{\pm\sqrt{(1/\hmo+1/4)/\eps}}$ in particular can
be significant.

Refs.\ \cite{CGRT} and \cite{GW3} showed that the canonically normalized 
radion field
is $\phi = f\hp$, where $f = \sqrt{6 M^3/k}$ is another scale of 
order $M_p$.\footnote{The choice $f=\sqrt{24 M^3/k}$ in ref.\ \cite{GW3}
seems to correspond to an unconventional normalization for $M_p$.}
The 4-D effective action for the radion and gravity is
\beqa
\label{rad_action}
	S &=& {M^3\over 2k} \int d^{\,4}x\, \sqrt{-g}\left(1-\hp^2\right) R
	\nonumber\\ 	
	&+& \int d^{\,4}x\, \sqrt{-g}\left( \sfrac12 f^2\partial_\mu\hp
	\partial^\mu\hp - V(\hp) \right),
\eeqa

\noindent
where $V(\hp)$ is the potential (\ref{eq:pot}) and $R$ is the Ricci
scalar.  To get the correct strength of gravity, we must therefore have
\beq
\label{Mpl}
	{M^3\over k} = M_p^2 = {1\over 8\pi G_{\sss N}}; 
	\qquad f = \sqrt{6} M_p.
\eeq
The radion mass
is $f^{-2}$ times the second derivative of $V$ evaluated at its minimum.
We find the value
\beq
\label{rad_mass}
	m^2_\phi = 4{\Lambda\over f^{2}}\left(
	1\!+\!\sfrac{\eps}{4}\!-\!\sfrac{\eps}{\hmo}\right)^{-1}\!
	\left(\sqrt{\phantom{AAAA}} \!\!\!\!\!\!\!\!\!\!\!\!\!\!\!\!\!\!
	1+\sfrac{4}{\hmo}
+2\sfrac{\sqrt{\eps}}{\hmo}\right)
	\,\eta^2\,
	\hp_+^2\, \eps^{3/2},
\eeq

\noindent which implies that $m_\phi$ is of typically of order $\eps^{3/4}$ times
the TeV scale.  The factor of $\eps^{3/4}$ leads to the prediction that
the radion will be lighter than the Kaluza-Klein excitations of the
graviton, which would also be a signal of the new brane physics
\cite{GW3}.  However, we see that the corrections due to finite $\hmo$,
which were not explicitly considered in \cite{GW3},
can possibly compensate this and make the radion heavier, if
$\hmo\sim\eps$.

Now let us turn to the evolution of $\hp$ in the early universe.  For
this purpose it is important to understand that the depth of the potential
at its global minimum, as well as the height of the bump separating
the minimum from $\hp=0$, is set by the TeV scale.  The values of the
potential at these extrema are approximately (to leading order in
$\eps_4$, but exact in $\eps_1$)
\beq
\label{exteq}
   V(\hp_\pm) \cong \mp 2\eta^2
	\Lambda \hp_\pm^4{\eps_4\sqrt{\eps_1+\eps_4}
	\over 1 + \eps_4-\eps_1}(1\pm\sqrt{\eps_1}).
\eeq
Since $\Lambda \sim M_p^4$, the depth at the minimum is 
$V(\hp_+)\sim -\eps^{3/2}O({\rm TeV})^4$---suppressed slightly by the
factor of $\eps^{3/2}$.  The height of the bump at $\hp_-$ can be
considerably smaller because of the exponential factors in 
(\ref{extrema}).  In fact 
\beq
	\left|{V(\hp_-)\over V(\hp_+)}\right| =
	\left({\hp_-\over\hp_+}\right)^4 \equiv \Omega^4
\eeq
where
\beq
\label{Omegaeq}
	\Omega \equiv \left({1\!-\!\sqrt{\eps_1\!+\!\eps_4}
	\over
	1\!+\!\sqrt{\eps_1\!+\!\eps_4} }\right)^{1/\eps} \sim\
	\exp\left(-\sqrt{\sfrac{1}{\eps}(1+\sfrac{4}{\hmo})}
	\,\right)
\eeq
For example, if $\eps = 0.01$ as suggested by \cite{GW}, $\Omega^4$
is less than $10^{-17}$.  If the brane potential parameter
$m_1$ is not large, so that $\hmo\lsim 1$, the 
suppression will be much greater. Figure 1 illustrates the flatness
of the potential for the case $\eps=0.2$, where the barrier is not
so suppressed.  The new mass scale $\Omega$ TeV $\ll$ 1 TeV is due
to the small curvature of the radion potential at the top of the barrier,
and its smallness will play an important role in the following.

Thus the barrier separating the true minimum at $\hp_+$ from the
false one at $\hp=0$ is extremely shallow.  Moreover a generic initial
condition for the radion is a value like $\hp\sim 1$, quite different
from the one we want to end up with, $\hp\sim 10^{-16}$.  Clearly, the
shape of the potential is such that, if we started with a generic
initial value for $\hp$, it would easily roll past the local minimum
and the barrier, hardly noticing their presence.  The point $\hp=0$
toward which it rolls is the limit of infinite brane separation,
phenomenologically disastrous since gravity no longer couples at all to
the visible brane in this limit.

One might wonder whether inflation could prevent this unwanted outcome,
since then there would be a damping term in the $\phi$ equation of motion,
possibly causing it to roll slowly:
\beq
	\ddot\phi + 3H\dot\phi + V_{\rm eff}'(\phi) = 0.
\eeq
Indeed, with sufficiently large Hubble rate $H$, the motion could 
be damped so that $\phi$ would roll to its global minimum.  The
condition for slow-roll is that
\beq
\label{sr}
	V_{\rm eff}'' \ll 9 H^2.
\eeq
However, inflation is a two-edged sword in this instance, because the
effective potential $V_{\rm eff}$ for the radion gets additional
contributions from the curvature of the universe during inflation.
From eq.\ (\ref{rad_action}) one can see that
\beqa
\label{Vcorr}
	V_{\rm eff}(\hp) &=& V(\hp) + { M^3\over 2k}R\hp^2\nonumber\\
			&=& V(\hp) + 6{ M^3\over k}H^2\hp^2\nonumber\\
			&=& V(\hp) + H^2\phi^2\,
\eeqa
using the relation $R = 12H^2$ which applies for de Sitter space,
and eq.\ (\ref{Mpl}).
The new term tends to destroy the nontrivial minimum of the radion
potential.  One can estimate the relative
shift in the position of the minimum as
\beq
	{\delta\hp_+\over\hp_+} = - {\delta V'(\hp_+)\over \hp_+
	V''(\hp_+)} = - {H^2\over m^2_\phi}
\eeq
This should be less than unity to avoid the disappearance of the 
minimum altogether.  

Combining the requirement that the global minimum survives with the
slow-roll condition (\ref{sr}), evaluated near the minimum of the
potential, we find the following constraint on the Hubble rate:
\beq
\label{Hcond}
	\sfrac19 m^2_\phi \ll H^2 < m^2_\phi.
\eeq
This is a narrow range, if it exists at all.  In fact, one never expects
such a large Hubble rate in the Randall-Sundrum scenario since the TeV
scale is the cutoff: $H$ should never exceed $T^2/M_p \sim 10^{-16}$
TeV if the classical equations are to be valid.  
 The problem of radion stabilization might also be exacerbated 
because contributions to the energy density of the
universe which cause inflation can give additional terms of the type
$\hp^2$ to $V_{\rm eff}$ which
are not considered in the above argument.  For example, 
\twocolumn[\hsize\textwidth\columnwidth\hsize\csname@twocolumnfalse%
\endcsname
\leftline{\epsfxsize=7.0in\epsfbox{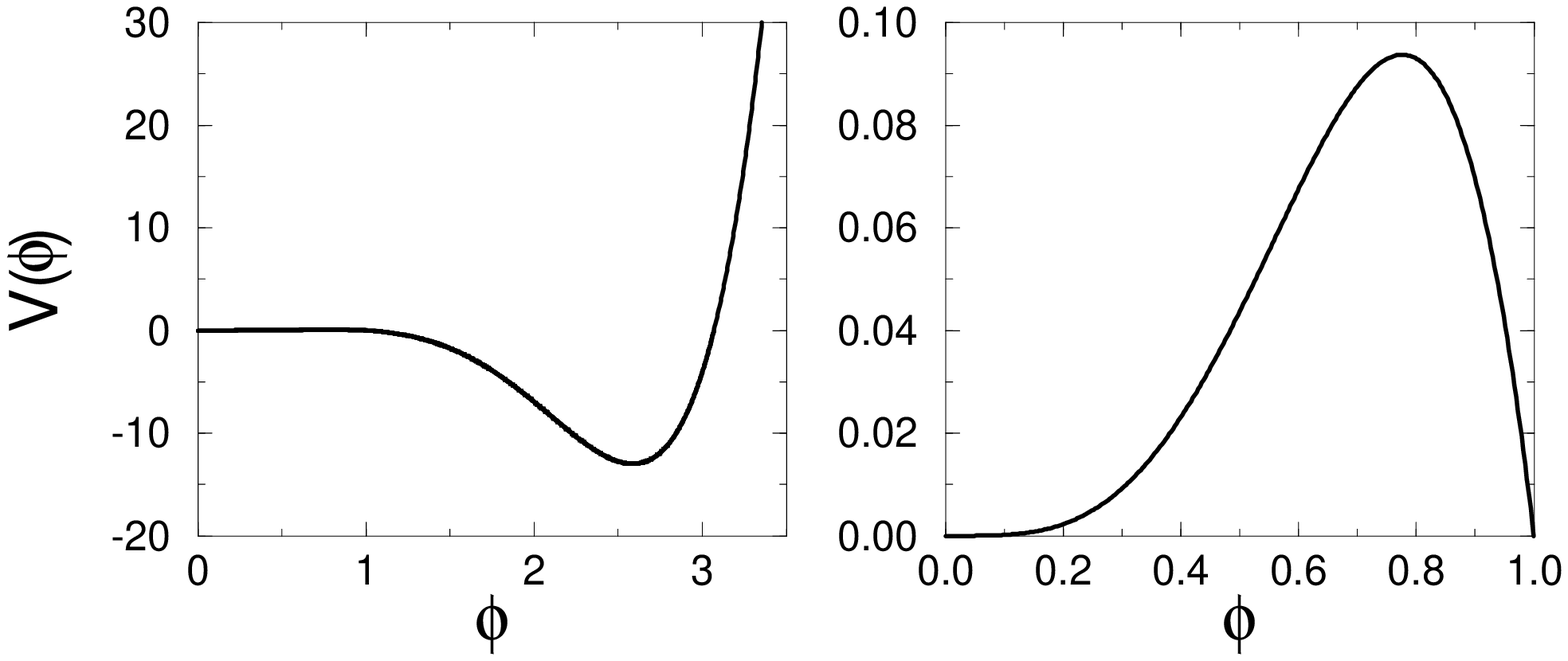}}
\vspace{0.in}
\noindent {\small 
Figure 1: $V(\phi)$ versus $\phi$ for $\eps =0.2$, showing the 
smallness of the barrier (right) relative to the minimum (left).
Notice the difference in vertical scales.}
\vspace{+0.1in}
]
\noindent 
a field in the bulk which does not have
its endpoints fixed on the branes has a 5-D energy density which is
peaked on the visible brane \cite{GW1}, $\rho_5(y)\sim \rho_0 e^{2kby}$, and
gives a contribution to $V_{\rm eff}$ of
\beqa
	\delta V &\sim& \rho_0 \int_0^1 dy\, b\, e^{-4kby}\rho_0 e^{2kby}
	\nonumber\\
	&=& b\rho_0 \hp^2,
\eeqa
remembering that $e^{-kb}=\hp$.  Such a contribution could destroy the
nontrivial minimum even if (\ref{Hcond}) is satisfied.  In any case, it
does not appear to be natural to tune the Hubble rate during inflation
to try to solve the stabilization problem.

\section{Phase Transition to the True Vacuum}

Since the barrier of the radion potential is too small to prevent the
radion from rolling into the false minimum, perhaps we can take
advantage of this smallness to get tunneling or thermal transitions
back into the true
vacuum.  The situation is quite similar to that of ``old inflation''
\cite{Guth}, except that in the latter, the transition was never able
to complete because the universe expanded too rapidly compared to the
rate of nucleation of bubbles of the true vacuum.  In the present case
this problem can be avoided because we are not trying to use the radion
for inflation.  Indeed, a small amount of inflation may take place
before the tunneling occurs, since the radion potential is greater than
zero at $\phi=0$, but we will not insist that this be sufficient to
solve the cosmological problems inflation is intended to
solve---otherwise we would be stuck with the problems of old
inflation.  Instead we will assume that inflation is driven by some
other field, and consider the transitions of the radion starting from the
time of reheating.  The criterion that the phase transition completes
is that the rate of bubble nucleation per unit volume, $\Gamma/V$,
exceeds the rate of expansion of the universe per Hubble volume:
\beq
	{\Gamma\over V} \gsim H^4
\eeq
The reason is that the bubbles expand at nearly the speed of light, so the
relevant volume is determined by the distance which light will
have traveled by a given time, which is of order $1/H$.

\subsection{The Euclidean Bounce}

To compute the nucleation rate $\Gamma/V$, one must construct the bounce
solution which is a saddle point of the Euclideanized action
\cite{Coleman}, in other words, with the sign of the potential
reversed. This is a critical bubble solution with the boundary
conditions
\beqa
\label{bbcs}
	\phi(r)|_{r=0} &=& \phi_0;\qquad \phi'(r)|_{r=0} = 0;\nonumber\\
	\phi(r)|_{r\to\infty} &=& 0; 
	\qquad \phantom{0}\phi'(r)|_{r\to\infty} = 0.
\eeqa
The value of $\phi_0$ which ensures the desired behavior as $r\to\infty$
cannot be computed analytically because the motion of the field is
damped by the term $\phi'/r$ in the equation of motion. We will
consider bubble nucleation at finite temperature in the high $T$ limit,
where the bounce solutions are three dimensional.
The equation of motion is
\beq
   {1\over r^{2}} \left(r^{2}\phi'\right)' = +V_{\rm eff}'(\phi),
\eeq
where now $V_{\rm eff}$ includes thermal corrections, which are much
larger than the $H^2\phi^2$ term considered in eq.\ (\ref{Vcorr}):

\twocolumn[\hsize\textwidth\columnwidth\hsize\csname@twocolumnfalse%
\endcsname
\centerline{\epsfxsize=6.0in\epsfbox{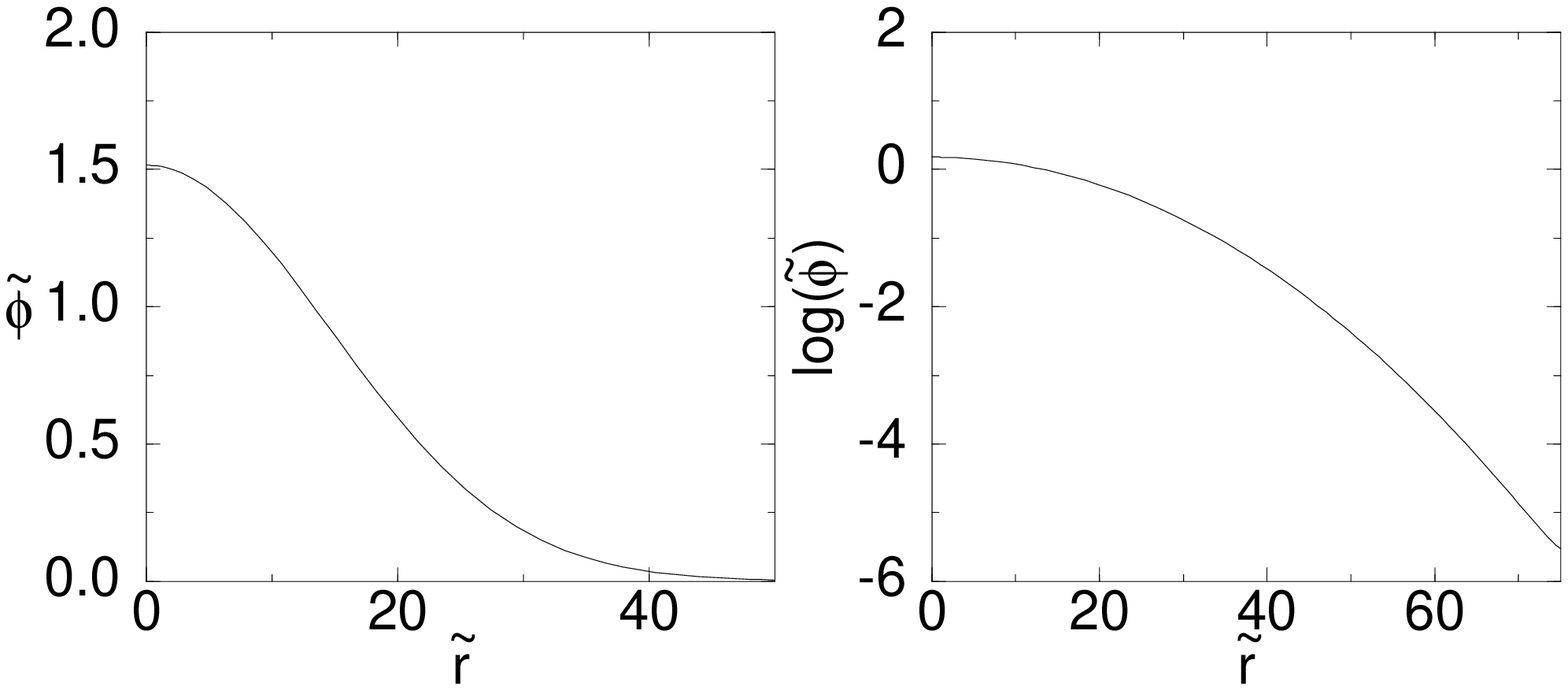}}
\vspace{0.in}
\noindent {\small 
Figure 2: The bounce solution, for the parameters $\eps=0.01$,
$\hmo = 0.1$, $m_\phi=T=100$ GeV and $M_{\rm TeV} = 1$ TeV.  
The rescaling $\phi\to\tilde\phi$ and
$r\to\tilde r$ is given in eqs.\ (\ref{rescale1}-\ref{rescale2}).}
\vspace{+0.1in}
]

\beqa
\label{Vcorr2}
	V_{\rm eff}(\phi) \cong V(\phi) &+&{T^2\over 24} m^2_{\rm th}(\phi)
	-{T\over 12\pi} m^3_{\rm th}(\phi) 
	\nonumber\\ &+& {c_b\over 64\pi^2}m^4_{\rm th}(\phi)
	 - V_0\\
	m^2_{\rm th}(\phi) =  V''(\phi) &+& \sfrac{1}{24}T^2 V^{(4)}(0)
\eeqa
where $c_b\cong 3.9$ if the renormalization scale is taken to be equal
to $T$.  The leading thermal correction is of order $T^2\phi^2$,
whereas $H^2\phi^2$ is suppressed by $T^2/M_p^2$ relative to this.  The
cubic term $m^3_{\rm th}$ becomes imaginary if $V''(\phi) +
\sfrac{T^2}{24}V^{(4)}(0)$ becomes negative; we take the real part
only.  The term $\sfrac{1}{24}T^2 V^{(4)}(0)$ represents the thermal
mass, which appears in the cubic and quartic terms when ring diagrams
are resummed \cite{Arnold}.  We subtract a constant term $V_0$ from $V_{\rm
eff}$ so that $V_{\rm eff}(0)=0$, as is needed to properly compute the
action associated with the bounce solution.

The thermal corrections to the effective radion potential cause the 
bounce solution to fall exponentially at large $r$:
\beq
\phi \sim  {c\over r}\, e^{-r/r_0},
\eeq
where $1/r_0 \sim \eta\sqrt{\Lambda/f^4}\,T$ if $\Lambda/f^4\ll 1$ [the
exact expression is $1/r_0 = \sqrt{\lambda U_*} T$, in terms of 
quantities to be defined below, in eqs.\ (\ref{rescale1}) and 
(\ref{Ustareq})].
Once the bounce solution is known, it must be substituted back into the
action, which can be written as 
\beq
S = {4\pi\over T}\int_0^\infty dr\,r^{2}
\left(\sfrac12\phi'^2 + V_{\rm eff}(\phi)\right).
\eeq
The nucleation rate is given by
\beq
\label{rate}
	{\Gamma\over V} = {|\omega_-|\over 2\pi}\left(S\over 2\pi\right)^{3/2}
	|{\cal D}|^{-1/2}\, e^{-S}
\eeq
where $\omega_-$ is the imaginary frequency of the unstable mode of
fluctuations around the the bounce solution, and 
${\cal D}$ is the
fluctuation determinant factor, to be described below.  A typical
profile for the bounce solution is shown in figure 2.

For the numerical determination of the bounce solution and action, 
as well as understanding their parametric dependences, it is
convenient to rescale the radius and the field by 
\beqa
\label{rescale1}	
   r&=& {\tilde r\over \sqrt{\lambda}T};\qquad 
\lambda = {\Lambda'\over f^4} c_-^2, \\
\label{rescale2}	
   \phi &=& Z\,T\tilde\phi;\qquad 
	Z = {fc_-^{1/\eps}\over T}.
\eeqa
Then the action takes the form
\beqa
\label{ba1}
S &=& {Z^2\over\sqrt{\lambda}}\,
\tilde S(\eps,\hmo,\lambda,Z);\\
\label{bounce_action}
\tilde S &=&  4\pi
\int_0^\infty \!\! d\tilde r\,\tilde r^{2}
\Biggl\{\sfrac12 \left( \tilde\phi'^2 + \tilde\phi^2 f_2 \right)
+ Z^2\tilde\phi^4 f_0\nonumber\\
&-& \frac{1}{12\pi}{\sqrt{\lambda}\over Z^2}\left[
	\left(\sfrac{c_+}{c_-} + 12 Z^2 \tilde\phi^2
	f_2 \right)^{3/2} - \left(\sfrac{c_+}{c_-}\right)^{3/2}
	\right] \nonumber\\
&+& {3 c_b\over 8\pi^2} \lambda f_2\tilde\phi^2 
\left(\sfrac{c_+}{c_-} + 6 Z^2 \tilde\phi^2 f_2 \right)\Biggr\}
\eeqa
where
\beqa
\label{feq}
	f_0(\tilde\phi) &=&  (\tilde\phi^\eps-1)
	(\tilde\phi^\eps-\sfrac{c_+}{c_-}) \nonumber\\
	f_{n+1}(\tilde\phi) &=& \left(1 +
	\sfrac{1}{4-n}\phi\partial_{\tilde\phi}
	\right) f_{n}(\tilde\phi)
\eeqa
In the following, it will be helpful to keep in mind that $Z$
can be extremely small, of order $\Omega$ in (\ref{Omegaeq}) when $\eps$ 
is small, whereas $\sqrt{\lambda}$ tends to be closer to unity,
depending on the mass of the radion and the definition of the TeV
scale (\ref{MTeVeq}):
\beq
	\sqrt{\lambda} = {\eps^{-3/4}\over 2\sqrt{6}}\, 
	{m_\phi\over M_{\rm TeV}}\,
{(1\!-\!\sqrt{\eps_1\!+\!\eps_4})\over
	\left(\sqe14
+2\sfrac{\sqrt{\eps}}{\hmo}\right)^{1/2} }
\eeq

In the rescaled variables, the value $\tilde\phi=1$ corresponds to the
first zero of the potential, which would be the starting point of the
bounce if energy was conserved in the mechanical analog problem, {\it
i.e.,} if there was no viscous damping term $\phi'/r$ in the equation
of motion.  The actual starting point turns out to have a value
in the range $\tilde\phi_0\sim 1.5- 3$ because of this.  
The rescaled action
$\tilde S$ depends mainly on the model parameters $\eps$ and $\hmo$, for 
the parameter values which are of interest to us.  
All the sensitive exponential dependence
on $\eps$, namely the factor $c_-^{1/\eps}$, is removed from $\tilde
S$.  Numerically we find that 
\beq
	\tilde S \cong \sfrac23 \hmo (\eps\hmo)^{-3/4},
\eeq
except when $\hmo$ becomes close to $\eps$.  For $\hmo$ slightly
smaller than $\eps$, $\eps_1$ starts to exceed 1, and $c_-$ becomes
negative, signaling the onset of an instability in the radion
potential toward coincidence of the two brane positions.  
Figure 3 shows the
dependence of $\log(\tilde S/\hmo)$ versus $\log(\eps\hmo)$.

\centerline{\epsfxsize=3.5in\epsfbox{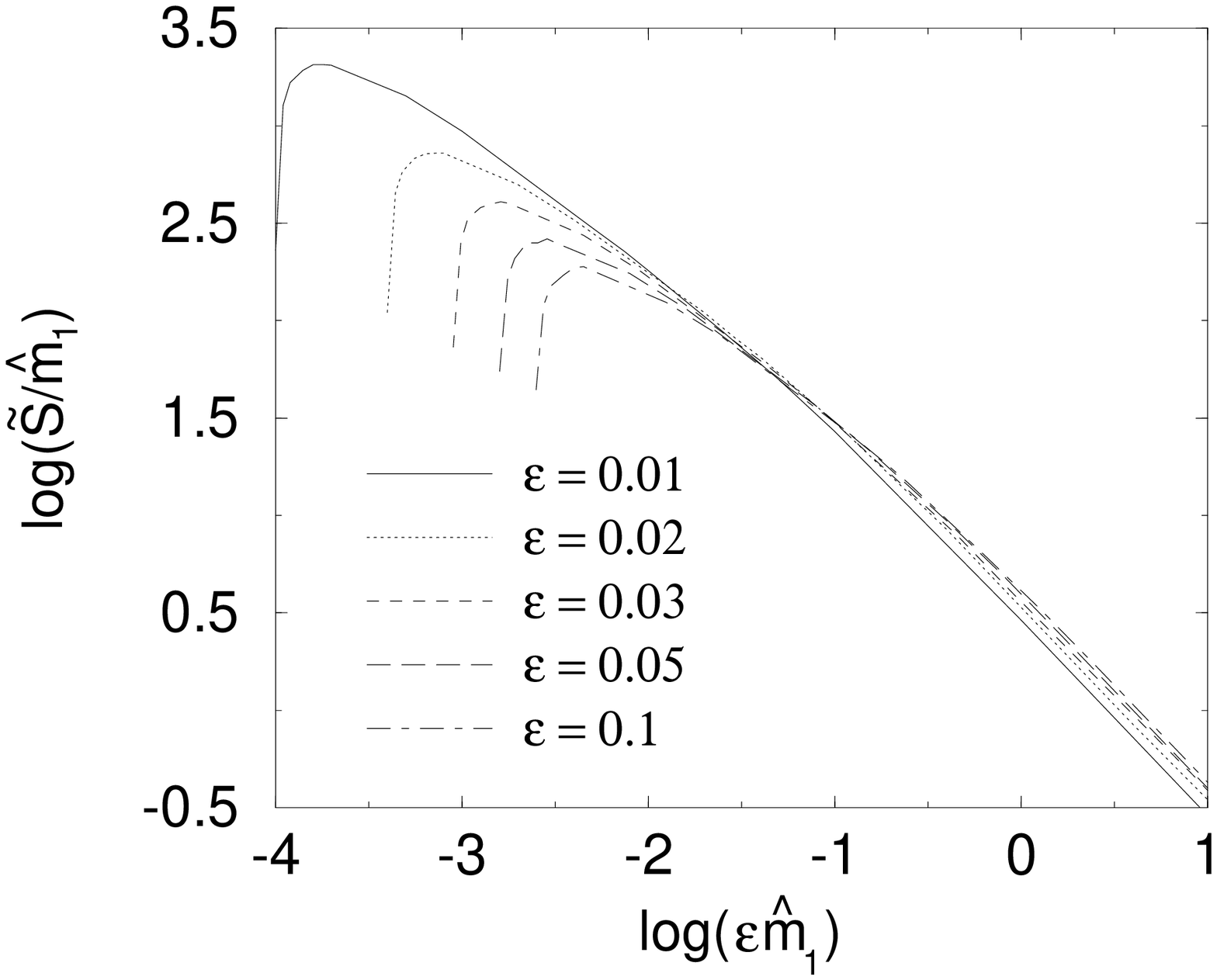}}
\vspace{-0.05in}
\noindent {\small 
Figure 3: $\log_{10}(\tilde S/\hmo)$ versus $\log_{10}(\eps\hmo)$, where
$\tilde S$ is the rescaled bounce action, eq.\ (\ref{bounce_action}).
The other parameters are $m_\phi=T=100$ GeV and $M_{\rm TeV} = 1$ TeV.}
\vspace{+0.1in}

We have computed the bounce solution and the corresponding action for a
range of parameters $\eps$ and $\hmo$ which can be consistent with the
solution to the hierarchy problem ({\it i.e.,} that $\hp \sim 10^{-16}$
at the global minimum).  The size of the bounce in position space,
measured as the width at half-maximum, is small near $\eps = \hmo$, and
reaches a larger constant value as $\hmo\to\infty$.  Using the rescaled
radial variable $\tilde r=r \sqrt{\lambda}T $, the dependence of the
width on $\eps$ and $\hmo$ is shown in figure 4.

\centerline{\epsfxsize=3.5in\epsfbox{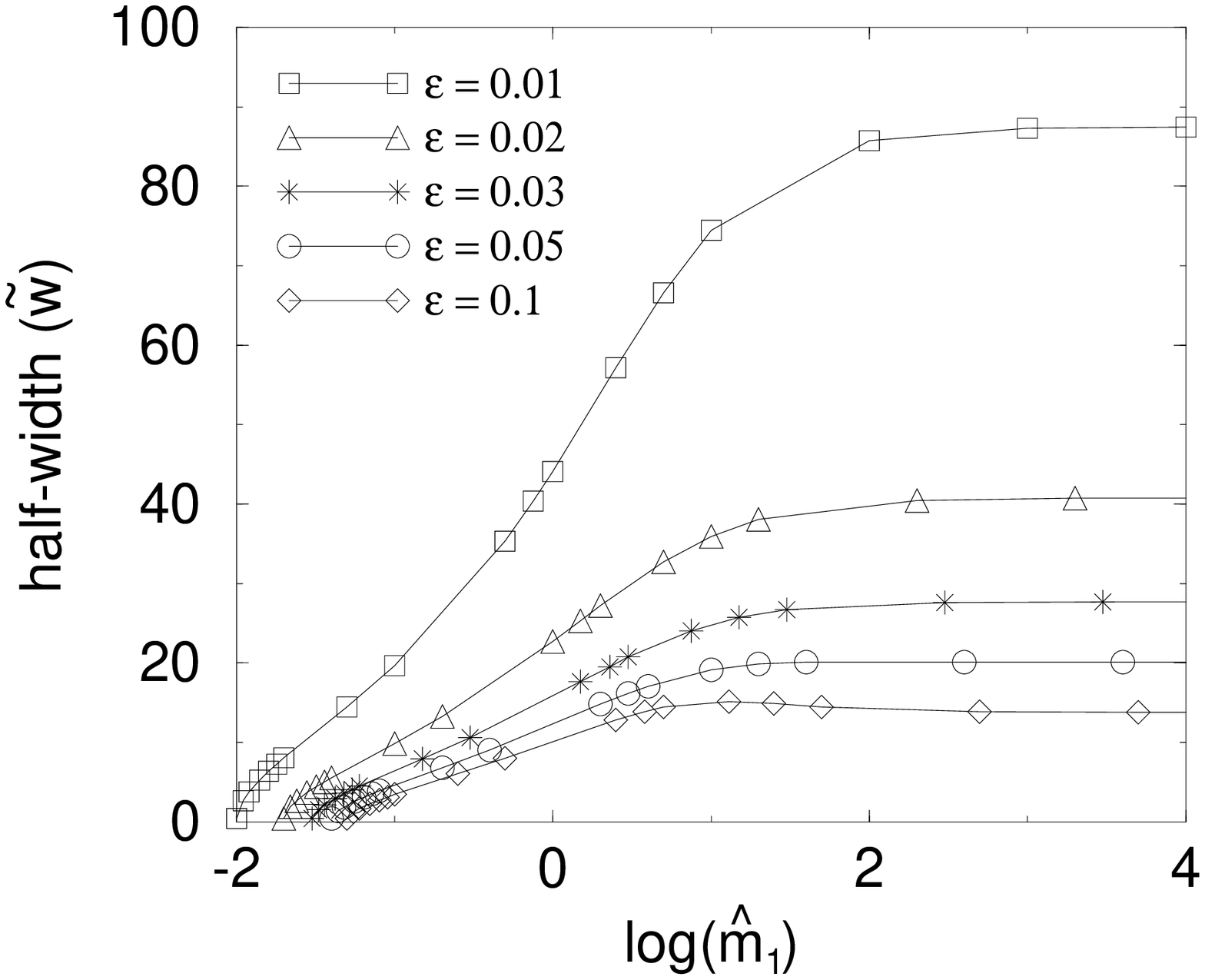}}
\noindent {\small 
Figure 4: Half-width $\tilde w$ of the bounce solution, in terms of the 
rescaled radial distance $\tilde r=r\sqrt{\lambda}T$, 
versus $\log_{10}(\hmo)$, for the same parameters as in figure 3.}
\vspace{+0.1in}

The action of the bounce can be much greater than or much less than 1,
depending on the parameters: for $\eps\sim \hmo \ll 1$, $S \ll 1$,
while for larger values of $\eps$ and $\hmo$, $S\gg 1$.  
Where the crossover occurs ($S\sim 1$) depends on $m_\phi$, $T$
and $M_{\rm TeV}$.  This behavior can be inferred from figure 3
(showing $\tilde S$) and the dependences of the coefficient in the
relation $S = (Z^2 /\sqrt{\lambda})\tilde S$.  Rather than
presenting further results for $S$ directly however, we will turn to
the more relevant quantity, the rate of bubble nucleation.  For this
we need to determine the prefactor of $e^{-S}$ in the rate.

\subsection{Prefactor of Bubble Nucleation Rate}

The bounce action is the most important quantity determining the rate
of tunneling, since it appears in the exponent of the rate
(\ref{rate}).  Since we do not have a model for the inflation and
reheating of the universe which must occur prior to the bubble
nucleation, hence an exact prediction for the reheating temperature
which enters the rate,  it would not be worthwhile to compute the
prefactors in eq.\ (\ref{rate}) very accurately; however we can
estimate their size.

First, consider the frequency $\omega_-$ of the unstable mode.  
$\omega^2_-$ is the negative
eigenvalue of the Schr\"odinger-like equation for small fluctuations
$\delta\phi$ around the bounce solution, which we will denote by
$\phi_b(r)$:
\beq
\label{fluceq}
	-\left(\delta\phi'' + {2\over r}\delta\phi'\right)
	+\left.{\partial^2 V_{\rm eff}\over\partial\phi^2}
	\right|_{\phi_b(r)}\, 
	\!\!\!\!\!\delta\phi = \omega^2_- \delta\phi
\eeq
Rescaling the radius and background field exactly as in
eqs.\ (\ref{rescale1}-\ref{rescale2}), eq.\ (\ref{fluceq}) becomes
\beqa	
\label{fluceq2}
	-\bigg(\delta\phi''  &+& \left. {2\over \tilde r}\delta\phi'\right)
	+ U(\tilde r)\, \delta\phi  = {\omega^2_-\over\lambda T^2} \,
	\delta\phi;
\eeqa

\twocolumn[\hsize\textwidth\columnwidth\hsize\csname@twocolumnfalse%
\endcsname
\centerline{\epsfxsize=6.0in\epsfbox{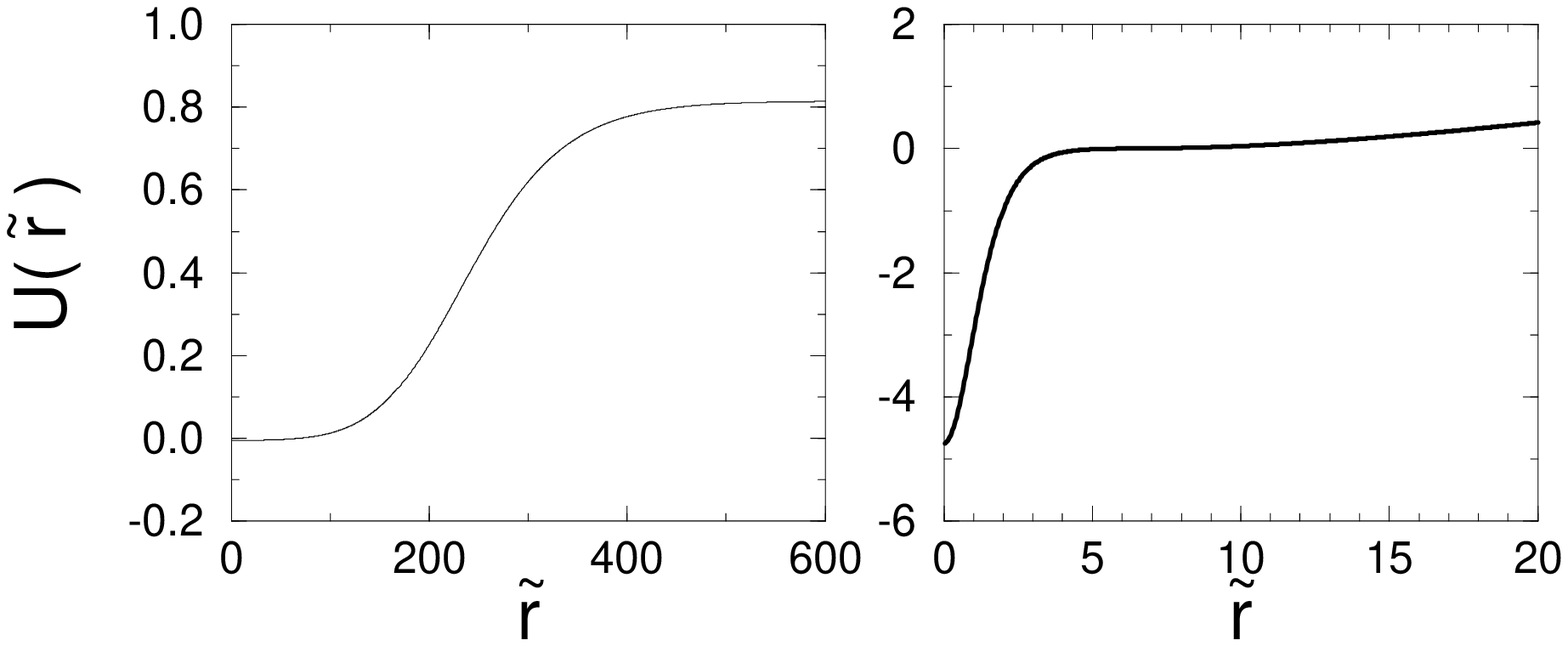}}
\vspace{0.in}
\noindent {\small 
Figure 5: The potential for small fluctuations around the bounce
solution, $U(\tilde r) = X^2\,V''(\phi_b(\tilde r))$,
as a function of $\tilde r$, for the parameters 
$\eps=0.02,$ $\hmo= 5$ (left) and $\eps = 0.1$, $\hmo = 25$ (right).  
}
\vspace{+0.1in}
]

\beqa
	U(\tilde r) &=& {1\over\lambda T^2} 
	{\partial^2 V_{\rm eff}\over\partial\phi^2}
	({\tilde\phi_b(r)}) \nonumber\\ 
	&=& f_4 + X - {3\sqrt{\lambda}\over\pi}\left(
	f_4\sqrt{\sfrac{c_+}{c_-}+X} + 
	{12 (Z\tilde\phi f_3)^2\over \sqrt{\sfrac{c_+}{c_-}+X}}
	\right)\nonumber\\
	&+& {6 c_b\over 8\pi^2}\lambda \left(\sfrac{c_+}{c_-}f_4+X(f_4+2f_3^2/f_2)
	\right)
\eeqa
where primes now denote $\sfrac{d}{d\tilde r}$, $X = 12 Z^2 \tilde\phi^2 f_2$,  
and the $f_i$ are defined in eq.\ (\ref{feq}).  
Except when the radion mass is signficantly larger than 100 GeV, $\lambda$
is much smaller than 1, and $U$ is dominated by the first two terms in
(\ref{fluceq2}).  Of these, the first ($f_4$) always dominates if $Z\ll 1$, 
while the
second ($X$) can be important near $r=0$ if $Z \gsim 1$.  The two different
cases are illustrated in figure 5.  Since $1 < \tilde\phi_0^\eps < 
\sfrac{c_+}{c_-}$, both $f_4$ and $X$ are negative at $r=0$, so that 
\beqa
	U_0 \equiv U(0) &\cong& -\eps^{3/2}\sqe14\nonumber\\
	&\times&\left(
	2 +\ln\tilde\phi_0 + 12 Z^2\tilde\phi_0^2(\sfrac12 + \ln\tilde\phi_0)
	\right)
\eeqa
and thus the smallest eigenvalue of eq.\ (\ref{fluceq}) is negative.
This is the unstable mode of the saddle point solution, with imaginary
frequency of order
\beq
\label{unstab_freq}
	\omega^2_- \sim  U_0 \lambda T^2.
\eeq
Recall that $|\omega_-|$ appears in the prefactor of the nucleation
rate $\Gamma/V$. 

As $\tilde r\to\infty$, $U(\tilde r)$ approaches a maximum value
\beq
\label{Ustareq}
	U_* \equiv U(\infty) = \sfrac{c_+}{c_-}\left(1-
	\sfrac{3}{\pi}\sqrt{\lambda\sfrac{c_+}{c_-}}
	+ \sfrac{3c_b}{4\pi^2}\lambda\sfrac{c_+}{c_-} \,\right),
\eeq
which determines the asymptotic behavior of the fluctuations
at large $\tilde r$:  $\delta\phi \sim e^{-\sqrt{U_0}\tilde r}$.
The fluctuations around the false vacuum state ($\phi=0$) thus have a mass 
given by
\beq
\label{Umass}
	m^2 = V''_{\phi=0} =  U_* \lambda T^2,
\eeq
which will be relevant for the following.

Next we must estimate the functional determinant factor,
\beq
	{\cal D} = {\det'(-\partial_\tau^2-\nabla^2 + V''(\phi_b))\over
	\det(-\partial_\tau^2-\nabla^2 + V''(0))},
\eeq
where $\tau$ is imaginary time ($\tau\in[0,1/T]$), $\nabla^2$ is
the three-dimensional Laplacian, $\phi_b$ is the bounce solution, and 
the prime on $\det'$ means that the three translational zero-mode eigenvalues must be
omitted from the determinant for fluctuations around the bounce.  These
zero modes correspond to spatial translations of the bubble solution.
Because of their removal, ${\cal D}$ has dimensions of (energy)$^{-6}$, 
as is required to get a rate per unit volume in eq.\ (\ref{rate}).

Ref.\ \cite{BK} has given a thorough account of how to compute ${\cal
D}$ by a method which was discussed for one-dimensional systems in
\cite{Coleman2}.  In 3-D one should classify the eigenvalues of the
fluctuation operators by the quantum numbers $n$ and $l$, denoting
Matsubara and angular momentum excitations, respectively.  Then ${\cal
D}$ can be written as a product, ${\cal D}=\prod_{n,l} {\cal
D}_{n,l}$.


Ref.\ \cite{BK} shows that the contribution to ${\cal D}$ from the
$l$th partial wave can be expressed, to leading order in a
perturbative expansion in the potential $U(r)$, as
\beq
	{\cal D}_{n,l} \cong \left(1 + h^{(1)}_l\right)^{2l+1}.
\eeq
The quantity $h^{(1)}_l$ has the Green's function solution
\beqa
\label{heq}
	h^{(1)}_l &=& 2 \int_0^\infty\!\! dr\, r\, I_{l+1/2}(\kappa r)
	\, K_l(\kappa r)\, \left( V''({\phi_b(r)})-m^2\right)\nonumber\\
 &=& 2 \int_0^\infty\!\! d\tilde r\, \tilde r\, I_{l+1/2}
	(\sfrac{\kappa \tilde r}{\sqrt{\lambda}T})
	\, K_l(\sfrac{\kappa \tilde r}{\sqrt{\lambda}T})\, 
	\left(U(\tilde r)-U_*\right)
\eeqa
using the modified Bessel functions $I$ and $K$, and the mass $m$
of the field in the false vacuum, eq.\ (\ref{Umass}). 
For general Matsubara frequencies, $\nu=2\pi nT$, one defines 
$\kappa = \sqrt{ m^2 + \nu^2}$.

The subdeterminant for the $n=0$ (zero-temperature) sector of the
theory has the usual ultraviolet divergences of quantum field theory,
namely the vacuum diagram $\bigcirc\!\!\bullet$ (the dot represents one
insertion of $V(\phi)$), which should be absorbed by renormalization of
the zero of energy for the radion potential.  Since we are not
attempting to solve the cosmological constant problem here, we are
going to ignore all of this and compute only the factor ${\cal
D}_{0,1}$, which contains the translational zero modes---or more
precisely, which has the zero modes removed.  This removal is
accomplished by replacing 
\beq
\label{zeromode}
	1+h^{(1)}_1\to {d  h_1^{(1)}\over d\kappa^2}
\eeq
Notice that this quantity has dimensions of (mass)$^{-2}$, and there are
$2l+1 = 3$ such factors, so that $|{\cal D}|^{-1/2}$ has dimensions of
(mass)$^{3}$, as required.  From eq.\ (\ref{heq}) one can show that
\beqa
 &&{d  h_1^{(1)}\over d\kappa^2}={1\over\lambda T^2 U_*^2}\, I_{\sss U};
\nonumber\\ 
&& I_{\sss U}
\equiv \int_0^\infty\!\!\!\! dy\,y^2 \left(I_{3/2}(y)K_1(y)\right)'
\left(U(\sfrac{y}{\sqrt{U_*}})-U_*\right).
\eeqa
We have numerically evaluated the integral $I_{\sss U}$ for each set of parameters.
Our estimate for the fluctuation determinant factor in the
nucleation rate can then be written as
\beq
|{\cal D}|^{-1/2} \sim \left({\lambda T^2 U_*^2\over  I_{\sss U} }\right)^{3/2}
\eeq

\subsection{Results for Nucleation Rate}

Putting the above ingredients together to find the rate of bubble
nucleation per unit volume, $\Gamma/V$, we see that the latter depends
on five undetermined parameters: $\eps$, $\hmo$, $m_\phi$, $M_{\rm TeV}$
and the temperature $T$.  Ref.\ \cite{CGRT} showed that, as long as the
energy density on the TeV brane is much less than $M_{\rm TeV}^4$,
the usual 4-D effective theory governs the Hubble rate:
\beq
\label{Fried}
	H^2 = {\rho\over 3 M_p^2},
\eeq
where $\rho$ is the total energy density,
\beqa
\label{eden}
	\rho &=& g_* {\pi^2\over 30} T^4 + \rho_\phi;\nonumber\\
	\rho_\phi &=& \sfrac12\dot\phi^2 + V(\phi)-V(\phi_+)\nonumber\\
	&\cong& \sfrac38 M^2_{\rm TeV} m^2_\phi { \sqe14
	(1+\sqrt{\eps_1}) \over \sqe14 + 
	2\sfrac{\sqrt{\eps}}{\hmo}}
\eeqa
We take the number of relativistic degrees of freedom, $g_*$, to be 100.
The kinetic energy of the radion is zero since $\phi=0$ in the 
false vacuum, so  $\rho_\phi$ is essentially the potential energy of the
radion in the false vacuum, assuming the 4-D cosmological
constant is zero: $V(\phi)-V(\phi_+)=|V(\hp_+)|$, which is given
by eq.\ (\ref{exteq}). Depending on the parameters, this can be
comparable in size or dominate over the energy density of radiation.
Using our estimates for the prefactor of the tunneling rate, 
the logarithm of the ratio of $\Gamma/V$ to $H^4$ can be written as
\beq
\label{lograte}
	\ln{\Gamma\over VH^4} \cong  \ln\left(
	{\lambda^2 \sqrt{U_0}\, U_*^3 \over (2\pi)^{5/2}}\,T^4
	\left({3M_p^2 \over \rho}\right)^2
	\left({S\over I_{\sss U}}\right)^{3/2} \right) - S
\eeq
where $S$ is the action of the bounce solution.
The criterion for completion of the
phase transition to the true vacuum state is that $\ln(\Gamma/VH^4)>0$.
The saddle point approximation leading to eq.\ (\ref{rate}) 
is only valid if the action $S$ is not much less
than 1.  Otherwise, the barrier is not effective for preventing the field
from rolling to the true minimum, as in a second order phase transition.
This situation occurs in the vicinity of $\ln(\Gamma/VH^4)\sim 150$ in
the following results; thus the transition region where $\Gamma/VH^4=1$
is well within the realm of validity of the approximation.

In figure 6 we show the contours of constant $\ln(\Gamma/VH^4)$ in the
plane of $\log_{10}(\hmo)$ and $\log_{10}(\eps)$, starting with the
fiducial values $T=m_\phi=100$ GeV, $M_{\rm TeV}= 1$ TeV for the other
parameters, and showing how the results change when any one of these
is increased.  The dependences can be understood from the prefactor
$Z^2/\sqrt{\lambda}$ in the action, eq.\ (\ref{ba1}):
\beq
	{Z^2\over \sqrt{\lambda}} \sim \eps^{3/4} {M_{\rm TeV}^3\, \Omega^2
	\over T^2\, m_\phi},
\eeq
where we recall that $Z$ and $\lambda$ are given by
(\ref{rescale1}-\ref{rescale2}) and $\Omega$ by (\ref{Omegaeq}).
The factor $\Omega$ is responsible for suppressing the bounce action
when $\eps\ll 1$ or $\eps\hmo\ll 1$, explaining the shape of the allowed
regions in each graph.  Nucleation of bubbles containing the true
minimum becomes faster when the temperature
or the radion mass is increased, but slower if the definition of the 
TeV scale in increased.   These dependences are dictated not only by
the size of the barrier between the two minima in the effective potential,
but also by the size of the bubbles.

Interestingly, the borderline between allowed and forbidden regions of
parameter space falls within the range which is relevant from the point
of view of building a model of radion stabilization.  That is, some choices
which would otherwise have been natural and acceptable are ruled out by
our considerations.  We see furthermore that the choice of $\hmo\to\infty$,
as was effectively focused on

\twocolumn[\hsize\textwidth\columnwidth\hsize\csname@twocolumnfalse%
\endcsname
\centerline{\epsfxsize=6.0in\epsfbox{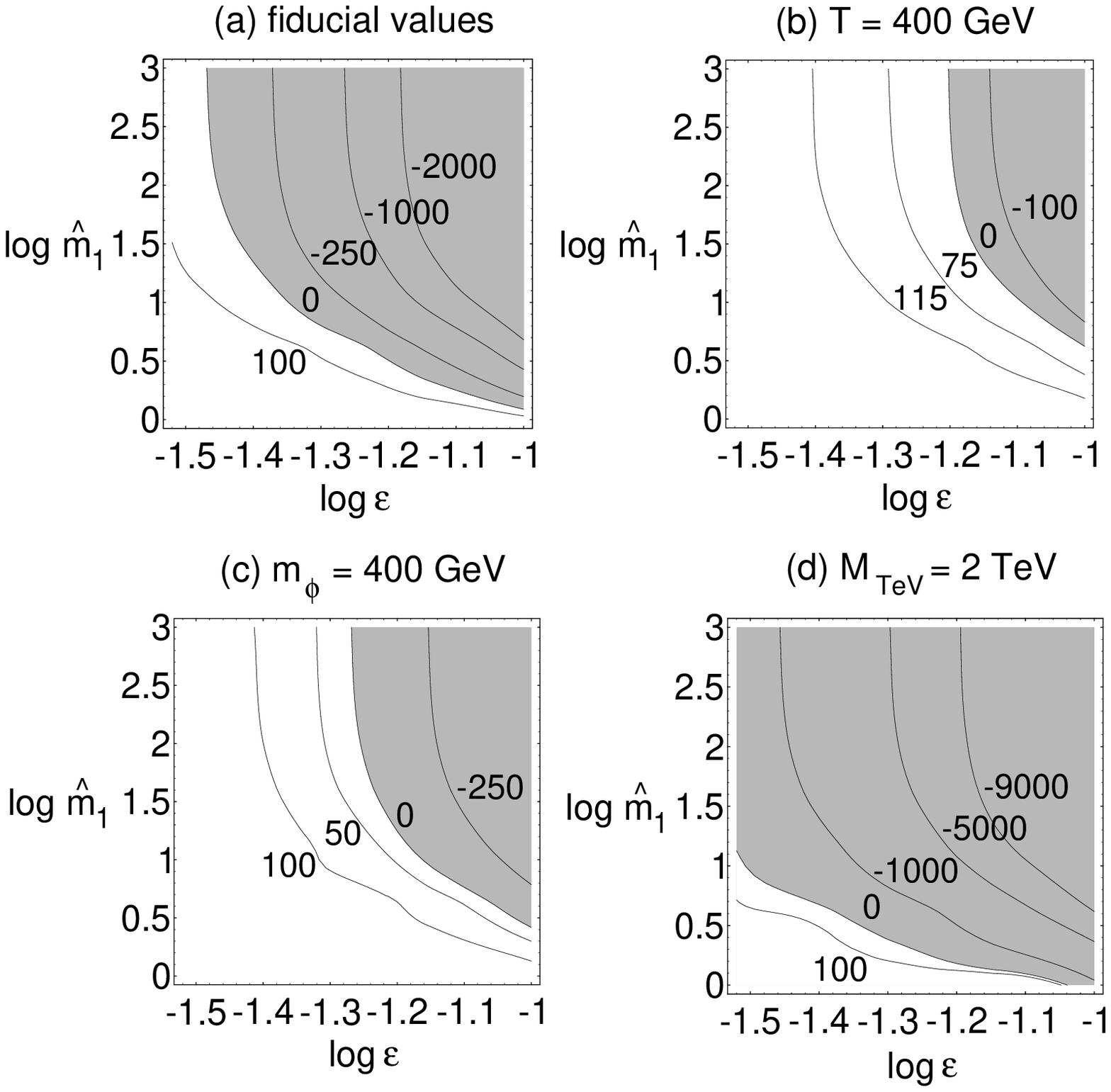}}
\vspace{0.4in}
\noindent {\small 
Figure 6: Contours of $\ln(\Gamma/VH^4)$ in the plane of $\log_{10}(\hmo)$
versus $\log_{10}(\eps)$.  The shaded regions are where the tunneling rate
is too small for the phase transition to complete.
Figure (a) has $T=m_\phi=100$ GeV, $M_{\rm TeV}
= 1$ TeV.  The other figures are the same except for the following changes:
(b) $T = 400$ GeV; (c) $m_\phi = 400$ GeV; (d) $M_{\rm TeV} = 2$ TeV.
}
\vspace{+0.4in}
]

\noindent in ref.\ \cite{GW}, is not the optimal one for achieving a large
nucleation rate. 

It might be thought that our analysis is rendered less important by the
fact that one can always obtain fast nucleation simply by going to
high enough temperatures.  However it must be remembered that the TeV scale
functions as the high-energy cutoff in the Randall-Sundrum scenario: the
whole semiclassical description breaks down at super-TeV scales, where
quantum gravity effects start to become important.  From this point of
view, the temperatures of $100-300$ GeV which we are discussing are 
already rather high, and a fairly efficient mechanism of
reheating at the end of inflation will be needed to generate them.

\section{Discussion}

In this paper we have presented a somewhat simpler model of radion
stabilization by a bulk field ($\psi$) than that of Goldberger and Wise
\cite{GW};  although the physics is qualitatively identical, we are able to
write the radion potential exactly, and thus explore the effect of letting
the stabilizing field's VEV's on the branes be pinned more or less strongly
to their minimum energy values.  One such effect is that the mass of the
radion can be significantly increased for small values of the parameter
$m_1$, which is the coefficient of the potential for $\psi$ on the TeV brane.
Moreover if $m_1/k\equiv \hmo$ is accidentally close to $\eps$, approximately
the minimum value consistent with a stable potential, the radion mass
can start to diverge, by the factor $(1-\sfrac{\eps}{4} -
\sfrac{\eps}{\hmo})^{-1/2}$.  This modifies somewhat the expectation
expressed
in ref.\ \cite{GW3} that the radion mass will be small relative to the TeV
scale, due to a factor of $\eps^{3/4}$.

Our main focus was on the problem that the radion potential has a local
minimum at infinite brane separation, and that the barrier between the
true and false minima is so small that for generic initial conditions,
one would expect the true minimum to be bypassed as the radion field rolls
through it.  We showed that for a large range of parameters, the
high-temperature phase transition to the true minimum is able to complete,
thus overcoming the problem.  There are however significant constraints on
the model parameters, and the initial temperature after inflation, to 
insure this successful outcome.

There remain some outstanding issues.  The form of the radion effective
potential is such that the field is able to reach $\phi=0$ in a finite amount
of time; yet $\phi=0$ represents infinite brane separation in the extra
dimension.  This paradoxical situation may be due to the assumption that the
stabilizing field, $\psi$, is always in its minimum energy configuration at
any given moment.  In reality $\psi$ must require a finite amount of time to
respond to changes in the radion.  Thus one should solve the coupled problem
for time-varying $\phi$ and $\psi$ to do better.  This is probably a
difficult problem, which we leave to future study. 

A related question is whether it is correct to treat thermal fluctuations of
the radion field $\phi$ analogously to a normal scalar field with values
in the range $(-\infty,\infty)$.  Since $\phi$ is related to the size of the
extra dimension by $\phi=f e^{-kb}$, its range is $[0,f]$.  We have not
studied what effect this might have on the thermal part of the effective
potential; instead we assumed that the usual treatment suffices.

Another approximation we made was to ignore the back-reaction of the
stabilizing field on the geometry.  Ref.\ \cite{DeWolfe} has given a method
of finding exact solutions to the coupled equations for the warp factor
$a(y)$ and the stabilizing field $\psi(y)$.  This method cannot be applied in
the present case because it works only for bulk scalar potentials with a
special form that, among other things, requires them to be unbounded from
below.\footnote{In \cite{DeWolfe} this is asserted to be allowed because of
special properties of anti-de Sitter space; however we believe that the real
reason the bulk potential can be unbounded from below is that the potentials
on the branes have the correct sign to prevent an instability.} Moreover,
since the method of \cite{DeWolfe} generates only static solutions to the
equation of motion, it cannot be used to deduce the radion potential, which
is a probe of the response of the geometry when it is perturbed away from a
static solution.  On the other hand, \cite{DeWolfe} does show that the
neglect of the back reaction is justified for the parameter values which most
closely resemble the Goldberger-Wise model. 

One might at first feel uneasy about using a 4-D effective description
of the problem when in reality our initial condition is a universe with
an infinitely large extra dimension.  In the Randall-Sundrum scenario,
however, this is justified because the graviton is trapped on the
transverse length scale of $1/k\sim 1/M_p$, rather than the size of the
5th dimension (see also \cite{RL}). Moreover the 4-D Friedmann equation
(\ref{Fried}) was shown by ref.\ \cite{CGRT} to be valid without
actually assuming the radion to be stabilized.\footnote{The extra
dimension is free to expand in this case, and the kinetic energy of the
radion simply appears as an additional contribution to the energy
density of the universe, as in eq.\ (\ref{Fried}-\ref{eden}).}
  Difficulties with the ``wrong'' rate of expansion ($H\propto \rho$
instead of $H\propto\sqrt{\rho}$ \cite{BDL}) arise only when one
fine-tunes the brane energy densities to prevent radion motion even in
the absence of stabilization, which we are not doing here.  In any
case, changing the form of the expansion rate would have a small effect
on our results since this alters only the logarithmic term in
(\ref{lograte}), not the overwhelmingly dominant term $S$.

The problem of shallow barriers in moduli potentials is not unique to
the Randall-Sundrum scenario, and a new idea for addressing it was
recently presented in ref.\ \cite{Stein}.  The coupling of the kinetic
terms of matter fields to the modulus can give the damping of the
modulus motion needed to make it settle in the true minimum in some
cases.  This effect might provide an alternative to the thermal
mechanism we have discussed here.

\bigskip

We thank Mark Wise and Rob Myers for enlightening discussions, Guy
Moore for helpful correspondence and Genevi\`eve Boisvert for
perceptive criticisms of the manuscript.  \vspace{-0.25in}


\end{document}